# Protons at the speed of sound: Specific biological signaling from physics

Bernhard Fichtl[1,2], Shamit Shrivastava[3], Matthias F. Schneider[3]

[1]University of Augsburg, Experimental Physics I, Augsburg, 86159, Germany

[2]Nanosystems Initiative Munich NIM, Schellingstr. 4, 80799 München, Germany

[3]Department of Mechanical Engineering, Boston University, Boston, MA, 02215, USA

## Abstract

Local changes in pH are known to significantly alter the state and activity of proteins and in particular enzymes. pH variations induced by pulses propagating along soft interfaces (e.g. the lipid bilayer) would therefore constitute an important pillar towards a new physical mechanism of biochemical regulation and biological signaling. Here we investigate the pH-induced physical perturbation of a lipid interface and the physiochemical nature of the subsequent acoustic propagation. Pulses are stimulated by local acidification of a lipid monolayer and propagate – in analogy to sound – at velocities controlled by the two-dimensional compressibility of the interface. With transient local pH changes of 0.6 units directly observed at the interface and velocities up to 1.4 m/s this represents hitherto the fastest protonic communication observed. Furthermore simultaneously propagating mechanical and electrical changes in the lipid interface up to 8 mN/m and 100 mV are detected, exposing the thermodynamic nature of these pulses.

Finally, these pulses are excitable only beyond a threshold for protonation, determined by the $pK_a$ of the lipid head groups. When combined with an enzymatic pH-optimum, the proposed communication can be very specific, thus providing a new physical basis for intra- and intercellular signaling via two-dimensional sound waves at interfaces.

## Introduction

Significant part of the cells organization origins from membranes. Its basic structure, the so called bilayer, is formed from lipids and incorporates various other biomolecules including enzymes. The common teleological explanation for the existence of such membranes is the regulation of in- and outflux [1]. This article, however, is inspired by the mindset that biological interfaces in general and lipid interfaces in particular are much more than a compartmental element of the living cell. They are fluctuating thermodynamic systems, which can change state and integrate local actions by propagating waves. We argue that they are responsive as well as receptive and actively involved in providing specificity for biochemical processes of the cell. The state diagrams of model lipid interfaces similar to a single leaflet of such a bilayer are conveniently recorded using the Langmuir technique [2,3]. It allows for a well-defined thermodynamic system where variables, like molecular area ($A$), lateral pressure ($\pi$), temperature ($T$), surface potential ($\psi$) and pH can be varied quasi-statically over a broad range. The resulting state diagrams present a clear picture of strong thermodynamic couplings observed in these systems, i.e. changes in mechanical properties are inevitably coupled to thermal, electrical and electromagnetic properties of the membrane [4–6]. Interestingly, similar couplings have been observed even during non-equilibrium dynamic processes, in particular, during two dimensional sound propagation in lipid monolayers [7].

Protons are known to excite gels, plant cells as well as neurons as has been shown in numerous studies [8–11]. They also play a pivotal role in various signaling pathways. Typically, the function of proteins depends on the surrounding pH with some proteins even exploiting pH gradients, for instance the enzyme ATP synthase [12,13]. However, while the effects of equilibrium state of the interface on local protonation kinetics and long range proton conduction via diffusion have been thoroughly investigated [14–20], non-equilibrium proton dynamics remains unexplored. In particular with the direct observation of acoustically propagating pulses in lipid interfaces recently published by some of us [21], pH-pulses seem reasonable and would offer an unprecedented explanation on biological communication and the orchestration of all the individual elements of a cell.

Indeed, here we show that *i)* acoustic pulses can be excited in lipid monolayers through local acidification of the interface, *ii)* that the excitation is specific and exhibits a local pH threshold and *iii)* that the resulting pulse reversibly changes the local pH of the interface. With propagation velocities of ~1 m/s, these pulses are orders of magnitudes faster than the lateral proton translocation at membrane interfaces [15,16] and represent hitherto the fastest "protonic communication" observed. Finally, we discuss the potential of these pulses as a new mechanism for intra- and intercellular biological signaling.

## Results & Discussions

The following section is divided into three parts: At first we will demonstrate that local acidification of lipid monolayers leads to acoustically propagating pressure pulses. Secondly, we will show that the excitation involves head group protonation and thus directly relates to the pK$_a$ of the lipid head group, which in turn opens the door for specific excitation. The third part will provide evidence for the coupling between pressure pulses and pH of the interface, enabling the local control of pH from remote. These findings will be supported by surface potential measurements, additionally revealing the simultaneous propagation of an electrical pulse. In the final, concluding part we will discuss the biological relevance of these results and propose a new model for specific biological signaling.

### Acidic excitation of acoustic waves

The addition of hydrochloric acid (gas) onto a DMPS monolayer results in a propagating change of lateral pressure [Setup see Fig. 1]. In Figure 2(a) a typical time plot of the lateral pressure signal π(t), following an excitation by hydrochloric acid gas, is shown. At first the pulse reaches sensor 1, resulting in a strong lateral pressure decrease of around 2.0 mN/m and a relaxation back to equilibrium. Around 0.25 seconds after the first sensor detects the pulse, it reaches sensor 2 with slightly damped amplitude (~ 45 %) considering the macroscopic distance. From the time delay of the pulse between sensor 1 and sensor 2 the propagation velocity $c$ can be calculated [Fig. 2(b)]. In the case of a sound pulse $c$ should depend on the lateral density $\rho_0$ and the adiabatic compressibility $\kappa_S$ of the material. In the linear case:

$$c \sim \sqrt{\frac{1}{\rho_0 \kappa_S}}$$

$\kappa_S$ is not directly accessible, but is usually well approximated by the isothermal compressibility [5,21,22], obtained from the inverse derivative of the DMPS isotherm:

$$\kappa_T = -\frac{1}{A}\frac{\partial A}{\partial \pi}\bigg|_T$$

In accordance with its mechanical susceptibility, $c$ increases to almost 0.7 m/s in the liquid expanded-phase, followed by a drop to 0.6 m/s in its phase transition regime. At pressures beyond the phase transition region, in the liquid-condensed state, the propagation speed rapidly increases up to 1.4 m/s at 30 mN/m. The correlation between the mechanical properties of the interface and the pulse velocities illustrates the acoustic foundation of these pulses [21,23].

Propagating pulses can also be evoked by other acids, e.g. acetic acid or nitric acid. This indicates the protonic nature of the excitation process, since the only common features between the acids are their dissociated protons [SI – S2]. Furthermore the excitation and propagation of pulses is not limited to DMPS, but also works with other lipids, e.g. DPPG (negatively charged) or DPPC (zwitterionic), demonstrating the universality of the observed phenomenon [SI – S3].

However, not every addition of acid leads to a propagative pulse. There exists a lower and an upper pH threshold for the excitation as described next.

## Subphase pH and specific/threshold excitation

If the subphase of the lipid monolayer is too acidic or too alkaline, no pulses can be excited [SI – S4]. In order to explain the pH bulk dependency of the excitation, the isothermal pH behavior of DMPS is studied [Fig. 3(a)]. The plateau region of the isotherms represents the phase transition of the lipids from the liquid-expanded to the liquid-condensed state. At high pH values ($\geq 7$) as well as at low pH values ($\leq 4$) the phase transition pressure $\pi_T$ changes only slightly with pH. In between these two regions, $\pi_T$ strongly depends on the pH of the subphase, leading to a sigmoidal $\pi_T$-$pH$-profile of the lipid. This behavior is well known for charged lipids [24,25] and due to the protonation of the lipid head group. From the first derivate of the curve $\left(\frac{\partial \pi}{\partial pH}\right)_T$, we obtain a pK$_a$-value of 5.4 for the carboxyl group of DMPS, in good agreement with literature [24,26].

The dependency of the excitation on the pH of the subphase can now be easily explained by the sigmoidal pK$_a$-profile. At high pH values the change in surface pH has to be large enough, in order to facilitate the protonation of the lipids and thereby a detectable propagative change in lateral pressure. If the monolayer is already fully protonated, as it is the case for low pH values, the addition of acid does not lead to propagating pulses anymore. Thus the dynamic response of the interface to a certain excitation $\left(\frac{\partial \pi}{\partial pH}\right)_S$, depends to a great degree on its chemical properties and environment, exhibiting a maximum near the pK$_a$ of the lipid monolayer. This threshold behavior of the interface introduces "specificity" in the excitation process and allows to control signal strength, which, as described below, opens up new possibilities for "specific communication".

So far we observed that close to the pK$_a$ of the monolayer, local pH changes inevitably lead to lateral pressure changes. Therefore the question arises, if the inverse relationship holds, too: Do propagating pressure pulses evoke pH changes $\left(\frac{\partial pH}{\partial \pi}\right)_S$ at the interface?

## Propagating pH-pulses

Lipid conjugated fluorescence probes provide a fast, noninvasive and effective method for measuring the local pH at a lipid interface [27]. The emission characteristics of these probes are sensitive to pH changes in its environment, especially near its pK$_a$. Importantly, in a lipid monolayer the emission intensity at a certain wavelength is also a function of surface pressure and thus cannot be interpreted in terms of only pH changes [6]. To quantify changes in the optical signal, one is better off measuring the ratio of intensities at two different wavelengths, eliminating the trouble of having to deal with absolute intensities.

Figure 3(b) shows the intensity ratio $I_R = I_{535nm}/I_{605nm}$ as a function of lateral pressure between 5 and 8 mN/m during isothermal compression at different buffer pH values of 6.5, 7 and 7.5, respectively. Clearly, the ratio reacts sensitive to pH changes but not to lateral pressure changes at the lipid interface. For a pH increase of one unit from pH 6.5 to pH 7.5 $I_R$ increases linearly from 2.0±0.1 to 2.6±0.1. Hence $I_R$ can be used as a measure for the local pH and allows for studying possible pH changes at the interface during a propagating pressure pulse. It is important to note, that this behavior is of course a fingerprint of the inherent phenomenology of the specific dye used and *cannot* be generalized. Indeed, without careful calibration in the proper environment a change in intensity cannot be converted into a change in pH. The transfer from bulk to interface for instance can change the dyes characteristics entirely.

In Figure 4(a) the time course of the pH at the dye during a propagating lateral pressure pulse is shown. Obviously, the two signals correlate (inversely) and based on the quasi-static coupling [Fig. 3(b)] a pH increase of approximately 0.6 units at the interface takes place. Subsequently the monolayer relaxes back to equilibrium where the pressure as well as the interfacial pH reacquire their former values.

In the same way as proton addition leads to condensation an expansion leads to the liberation of protons from the interface and hence an increase in local pH [25]. Thus, the negative correlation between pressure and pH origins from the fact, that the propagating front is actually an expansion caused by the local acidification at the excitation site (15 cm distance from the point of detection). These observations present strong evidence of a decoupling between the interface and the bulk for pulse propagation: While during quasi-static *isothermal* compression the bulk can efficiently buffer the interface [Fig. 3(b)], during an *adiabatic* propagation the interface is rather unaffected by the bulk.

Further support of the proposed mechanism arises from surface potential measurements. Proton concentration at the interface $[H_I^+]$ and proton concentration in the bulk $[H_B^+]$ are related by a Boltzmann factor and thus exponentially depend on $\psi$ [25]:

$$[H_I^+] = [H_B^+] \exp(\frac{e\psi}{kT})$$

Hence, for a pH change of $\Delta pH \sim 0.6$ units at the monolayer we expect a surface potential change of $\Delta\psi \sim 35$ mV in the quasi-static case.

In Fig. 4(b) the simultaneously propagating variations in surface potential and lateral pressure during a pulse are shown. Clearly, electrical and mechanical response of the monolayer are coupled and for the given lateral pressure variation of ~2.4 mN/m, the surface potential changes by about 25 mV. This measured value for Δ$\psi$ agrees well with our theoretical prediction, considering its quasi-static limitation. Further it supports the proposed mechanism, since a local pH increase at the lipid monolayer should lead to a decrease in surface potential[28].

For stronger excitations we could record pulses with amplitudes >100 mV [SI – S5], reaching the order of action potentials observed in biological systems. We believe that this observation should be considered when engaging into the current controversial discussion on the underlying mechanism of action potentials[22,29]. Due to the high buffer concentrations used (debye length ~ 1 nm), $\psi$ rapidly decays, leading to electric field variations during the pulse of the order of $10^5$ V/m across the interface. It appears hard to imagine, that proteins with dipole moments or even charged amino acids are unaffected by these enormous propagating fields.

## Conclusion

We have shown that lipid monolayers enable propagating mechano-chemical-electrical pulses with velocities controlled by the compressibility of the monolayer. The acoustic waves can be evoked only above a certain threshold. This threshold is determined by the head group protonation and hence the p$K_a$ of the lipid interface.

It is important to point out, that the propagation of local pH perturbations as described follows from fundamental physical principles applied to the phenomenology of interfaces. Providing significant localized proton release, they have therefore to be expected to exist in biology as well, even if velocities and/or amplitudes may vary significantly. Previously we have therefore proposed acoustic pulses as a new physical foundation for biological communication [21,22]. The results presented, however, constitute a big leap as they *i)* bridge the gap between physics (adiabatic pulses), biology (lipid interfaces) and chemistry (local pH) and *ii)* introduce a thermodynamic concept of *specificity* [Fig. 5]. In the following we are taking the liberty to briefly outline our ideas. We imagine an (membrane-bound) enzyme, which – as we will explain - will first serve as stimulus and in the next step as receptor for specific pulses:

*Specific excitation*: In its catalytically active state many enzymes (e.g. esterases, lipases) will locally liberate protons [30,31]. If the proton concentration and hence the pH reaches a certain threshold and if the protonation of the lipids proceeds fast enough, a propagating sound pulse will be triggered [Fig. 5(a)]. Its amplitude depends on the $pK_a$ of the interface and on the strength of the excitation. Specificity comes in twofold: $pK_a$ and the inherent pH-optimum of the enzyme.

*Specific Interaction*: Due to its mechanical, electrical and in particular chemical properties, the propagating pulses will affect proteins at the interface (e.g. enzymatic activity). In Fig. 5(b) two possible interactions of a pH-pulse with an enzyme are shown: If the local pH ($pH_{loc}$) is far from the enzyme's pH optimum ($pH_{opt}$), the pH-pulse will have only minor impact on the enzyme's activity. If, however, the surrounding pH is close to the $pH_{opt}$ of the enzyme, the enzyme activity could change enormously: Increasing, if the local pH is shifted towards $pH_{opt}$ or decreasing when the local-pH is shifted away, i.e. ($pH_{loc}$- $pH_{opt}$) decreasing or increasing, respectively.

Taken together, only if *i)* the stimulus of enzyme A leads to a propagating pulse across the interface and *ii)* the pulse shifts the local pH at enzyme B towards or away from its $pH_{opt}$, effective and specific communication between enzyme A and B will take place. In this case specificity arises from two (nonlinear) transitions and thus from physical principles rather than structural considerations. Clearly, specificity can be further enhanced if nonlinear relations between activity and other physical parameters, e.g. compressibility, heat capacity, electrical capacity etc. exist. Such relations have indeed been observed extensively and the maximal activity of phospholipase $A_2$ and phospholipase C at the lipid phase transitions are two of the best studied examples [31–33].

Indications for the validity of our idea of specific communication can be obtained by considering the dye as a first order approximation for membrane proteins. The emission spectrum of a dye reflects the microscopic energy distribution that corresponds to the conformational space of the emission dipole and the associated layers of solvent molecules. For example the fluorescence moiety of the dye used in this study (Oregon green 488 conjugated to lipid DHPE) has a pKa of 4.7, which determines the protonation state of the dye with respect to the local pH. In general the charged [D-] form has a greater transition dipole which results in a stronger coupling with the solvation shell. While a greater transition dipole results in stronger emission intensity, a stronger solvation results in a redshifted spectrum both of which are clearly evident in pH dependence of the emission spectrum of Oregon green 488 [27]. Thus a change in emission spectrum of the dye during a pulse as shown here not only indicates a change in the protonation state of the dye but also a change in the energy landscape of the dye-solvent complex. Similar, for a significant pH change one should expect alterations in the conformational landscape of the enzyme-substrate complex, e.g. by a change of the protonation state of the active site, which would affect the kinetics of the catalyzed reaction.

Taken into account all these factors, acoustic pulses represent an effective (fast), efficient (adiabatic) and specific tool to orchestrate the individual elements of a cell as well as to communicate between cells. It will be thrilling to see whether the type of communication suggested can be verified by single enzyme experiments along the lines of those [34–37].

# Methods

1,2-dimyristoyl-sn-glycero-3-phospho-L-serine (DMPS), 1,2-dihexadecanoyl-sn-glycero-3-phospho-(1'-rac-glycerol) (DPPG) and 1,2-dipalmitoyl-sn-glycero-3-phosphocholine (DPPC) were purchased from Avanti Polar Lipid (USA) and used without further purification. Monolayers were spread from a chloroform/methanol/water solution on a customary Langmuir trough (NIMA) until the desired lateral pressure was achieved. Measurements were started 10 minutes after solvent evaporation. If not further specified, all measurements were performed at 25°C on a buffer solution (pH 7.0) containing deionized water (resistivity >18 MΩcm), 100 mM sodium chloride, 10 mM phosphate buffer.

The Langmuir trough is equipped with two Wilhelmy plate pressure sensors, situated 15 cm apart from each other and a Kelvin probe sensor, facing pressure sensor 1 [Fig. 1]. The rapid readout of the sensors (10000 samples/s, 0.01 mN/m and 5 mV resolution) ensures accurate velocity and surface potential measurements. For the detection of fluorescent signals the Kelvin probe is substituted by an optical setup (not shown).

Pulses are excited blowing a fixed amount of pure nitrogen gas (5 ml for sole lateral pressure measurements and 25 ml for surface potential and pH measurements) through the gas phase of a glass bottle filled with 32% hydrochloric acid solution (for reference measurements: 100% acetic acid). pH measurements show that the excitation by 25 ml nitrogen gas drags along $(2.0 \pm 0.2) * 10^{-6}$ mol of hydrochloric acid. The acid/nitrogen gas mixture is then gently blown onto the lipid monolayer in order to prevent capillary waves. The excitation takes place 10 cm away from pressure sensor 1. The gaseous excitation allows for protonating bigger areas of the monolayer while using less amounts of acid than it would be possible for pipetting.

To exclude artifacts, we performed reference measurements on pure water surfaces. Neither nitrogen, nor hydrochloric acid or acetic acid showed any detectable pressure change at the surface. Furthermore pure nitrogen gas was blown onto a DMPS monolayer, to exclude any excitatory effect by $N_2$ [SI – S1].

pH changes at the interface are detected using lipid conjugated pH sensitive dye Oregon Green® 488 1,2-Dihexadecanoyl-sn-Glycero-3-Phosphoethanolamine spread along with DMPS (1 mol % dye). The emission of the dyes embedded in the monolayer was measured at 535 nm and 605 nm simultaneously with lateral pressure. Propagating changes were measured at a distance of 10 cm from the excitation spot. In order to rule out diffusion effects, a Teflon ring with a small opening facing away from the excitation spot was used to encircle the lipid monolayer around the spot for optical measurements.

## Acknowledgement

We are very grateful to Prof. Achim Wixforth and his chair (Experimental Physics 1 - University of Augsburg) for the support of this project. M.F.S. thanks Dr. Konrad Kaufmann (Göttingen), who inspired him to work in this field and introduced him to the thermodynamic origin of the phenomena and Einstein's approach to statistical physics. We also thank him for numerous seminars and discussions.

Financial support by BU-ME is acknowledged. M.F.S. appreciates funds for guest professorship from the German research foundation (DFG), SHENC-research unit FOR 1543. B. F. is grateful to Studienstiftung des deutschen Volkes for funding and to NIM for financial travel support.

# References

1. Alberts, B. & Wilson, J. *Molecular biology of the cell: [media DVD-ROM inside]*. (Garland Science, 2008).

2. Gaines, G. L. *Insoluble monolayers at liquid-gas interfaces*. (Interscience Publ., 1966).

3. Albrecht, O., Gruler, H. & Sackmann, E. Polymorphism of phospholipid monolayers. *J. Phys. Fr.* **39,** 301–313 (1978).

4. Heimburg, T. Mechanical aspects of membrane thermodynamics. Estimation of the mechanical properties of lipid membranes close to the chain melting transition from calorimetry. *Biochim. Biophys. Acta* **1415,** 147–162 (1998).

5. Steppich, D. *et al.* Thermomechanic-electrical coupling in phospholipid monolayers near the critical point. *Phys. Rev. E* **81,** 61123 (2010).

6. Shrivastava, S. & Schneider, M. F. Opto-Mechanical Coupling in Interfaces under Static and Propagative Conditions and Its Biological Implications. *PLoS One* **8,** e67524 (2013).

7. Griesbauer, J., Bössinger, S., Wixforth, A. & Schneider, M. F. Simultaneously propagating voltage and pressure pulses in lipid monolayers of pork brain and synthetic lipids. *Phys. Rev. E* **86,** 61909 (2012).

8. Fillafer, C. & Schneider, M. F. On the excitation of action potentials by protons and its potential implications for cholinergic transmission. *ArXiv e-prints* 1–15 (2014).

9. Krishtal, O. A. & Pidoplichko, V. I. A receptor for protons in the nerve cell membrane. *Neuroscience* **5,** 2325–2327 (1980).

10. Frederickson, R. C. A., Jordan, L. M. & Phillis, J. W. The action of noradrenaline on cortical neurons: effects of pH. *Brain Res.* **35,** 556–560 (1971).

11. Walters, D. H., Kuhn, W. & Kuhn, H. J. Action Potential with an Artificial pH-Muscle. *Nature* **189,** 381–383 (1961).

12. Fersht, A. *Structure and mechanism in protein science: A guide to enzyme catalysis and protein folding*. (Freeman).

13. Mitchell, P. Coupling of Phosphorylation to Electron and Hydrogen Transfer by a Chemi-Osmotic type of Mechanism. *Nature* **191,** 144–148 (1961).

14. Gabriel, B. & Teissié, J. Proton long-range migration along protein monolayers and its consequences on membrane coupling. *Proc. Natl. Acad. Sci. U. S. A.* **93,** 14521–14525 (1996).

15. Teissié, J., Prats, M., Soucaille, P. & Tocanne, J. F. Evidence for conduction of protons along the interface between water and a polar lipid monolayer. *Proc. Natl. Acad. Sci. U. S. A.* **32,** 3217–3221 (1985).

16. Serowy, S. *et al.* Structural Proton Diffusion along Lipid Bilayers. *Biophys. J.* **84,** 1031–1037 (2002).

17. Zhang, C. *et al.* Water at hydrophobic interfaces delays proton surface-to-bulk transfer and provides a pathway for lateral proton diffusion. *Proc. Natl. Acad. Sci. U. S. A.* **109,** 9744–9749 (2012).

18. Medvedev, E. S. & Stuchebrukhov, A. A. Mechanism of long-range proton translocation along biological membranes. *FEBS Lett.* **587,** 345–349 (2012).

19. Brändén, M., Sandén, T., Brzezinski, P. & Widengren, J. Localized proton microcircuits at the biological membrane–water interface. *Proc. Natl. Acad. Sci.* **103,** 19766–19770 (2006).

20. Sandén, T., Salomonsson, L., Brzezinski, P. & Widengren, J. Surface-coupled proton exchange of a membrane-bound proton acceptor. *Proc. Natl. Acad. Sci.* **107 ,** 4129–4134 (2010).

21. Griesbauer, J., Bössinger, S., Wixforth, A. & Schneider, M. F. Propagation of 2D Pressure Pulses in Lipid Monolayers and Its Possible Implications for Biology. *Phys. Rev. Lett.* **108,** 198103 (2012).

22. Shrivastava, S. & Schneider, M. F. Evidence for two-dimensional solitary sound waves in a lipid controlled interface and its implications for biological signalling. *J. R. Soc. Interface* **11,** (2014).

23. Griesbauer, J., Wixforth, a & Schneider, M. F. Wave propagation in lipid monolayers. *Biophys. J.* **97,** 2710–2716 (2009).

24. Marsh, D. *Handbook of lipid bilayers*. (CRC Press Taylor & Francis Group, 2013).

25. Träuble, H. in *Structure of Biological Membranes* (eds. Abrahamsson, S. & Pascher, I.) **34,** 509–550 (Springer US, 1977).

26. Cevc, G., Watts, A. & Marsh, D. Titration of the phase transition of phosphatidylserine bilayer membranes. Effects of pH, surface electrostatics, ion binding, and head-group hydration. *Biochemistry* **20,** 4955–4965 (1981).

27. Demchenko, A. P., Mély, Y., Duportail, G. & Klymchenko, A. S. Monitoring biophysical properties of lipid membranes by environment-sensitive fluorescent probes. *Biophys. J.* **96,** 3461–3470 (2009).

28. Vogel, V. & Möbius, D. Local surface potentials and electric dipole moments of lipid monolayers: Contributions of the water/lipid and the lipid/air interfaces. *J. Colloid Interface Sci.* **126,** 408–420 (1988).

29. Heimburg, T. & Jackson, A. D. On soliton propagation in biomembranes and nerves. *Proc. Natl. Acad. Sci. United States Am.* **102 ,** 9790–9795 (2005).

30. Silman, H. I. & Karlin, A. Effect of local pH changes caused by substrate hydrolysis on the activity of membrane-bound acetylcholinesterase. *Proc. Natl. Acad. Sci. U. S. A.* **58,** 1664–1668 (1967).

31. Holopainen, J. M., Angelova, M. I., Soderlund, T. & Kinnunen, P. K. Macroscopic consequences of the action of phospholipase C on giant unilamellar liposomes. *Biophys. J.* **83,** 932–943 (2002).

32. Op Den Kamp, J. A. F., De Gier, J. & van Deenen, L. L. M. Hydrolysis of phosphatidylcholine liposomes by pancreatic phospholipase A2 at the transition temperature. *Biochim. Biophys. Acta* **345,** 253–256 (1974).

33. Honger, T., Jorgensen, K., Biltonen, R. L. & Mouritsen, O. G. Systematic relationship between phospholipase A2 activity and dynamic lipid bilayer microheterogeneity. *Biochemistry* **35,** 9003–9006 (1996).

34. Lu, H. P., Xun, L. & Xie, X. S. Single-Molecule Enzymatic Dynamics. *Science (80-. ).* **282,** 1877–1882 (1998).

35. Xie, X. S. Enzyme Kinetics, Past and Present. *Science (80-. ).* **342,** 1457–1459 (2013).

36. Edman, L., Földes-Papp, Z., Wennmalm, S. & Rigler, R. The fluctuating enzyme: a single molecule approach. *Chem. Phys.* **247,** 11–22 (1999).

37. Gräslund, A., Rigler, R. & Widengren, J. Single Molecule Spectroscopy in Chemistry, Physics and Biology: Nobel Symposium. (2010).

# Figures & Figure Legends

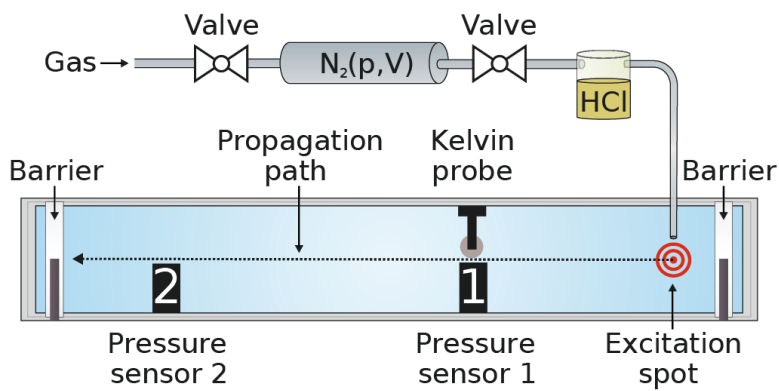

**Fig. 1:** The film balance setup (Langmuir trough) for analyzing propagating monolayer pulses consists of two pressure sensors and a Kelvin probe in order to measure mechanical and electrical changes at the lipid interface. In a typical experiment a fixed amount of nitrogen is blown through a glass bottle partly filled with an acid solution (in this case 32% HCl). The resulting gas mixture is then gently blown onto the lipid monolayer (red spot). Lateral pressure and surface potential are recorded and velocities are calculated. Two moveable Teflon barriers enable us to compress or expand the lipid film and thereby to record lateral pressure and surface potential isotherms. For fluorescent pH-measurements the Kelvin probe is exchanged by an optical setup (not shown). The dyes are excited at 465 nm and the emission is measured at 535 nm and 605 nm.

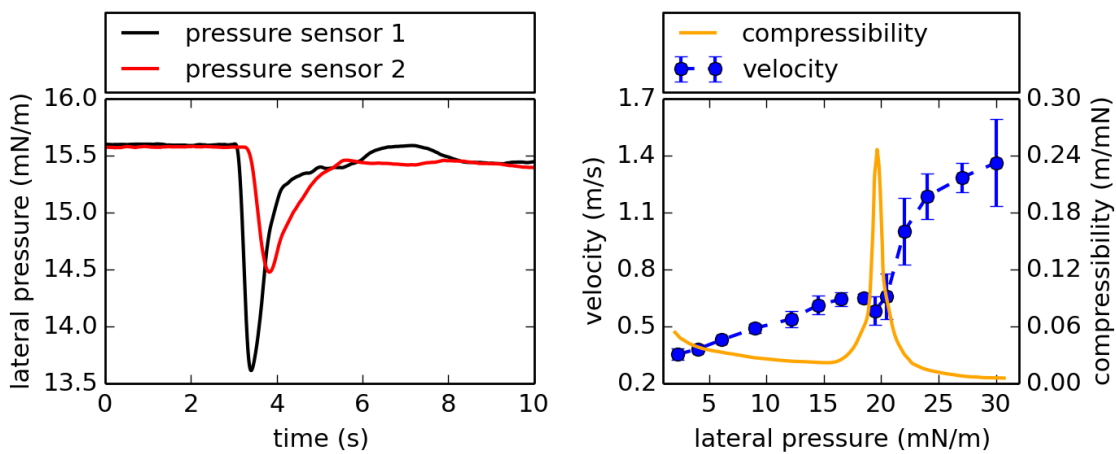

**Fig. 2:** Acidic excitation of acoustic waves: **(a)** Time course of a typical lateral pressure pulse traveling from sensor 1 to sensor 2 in a DMPS monolayer excited by hydrochloric acid. The amplitude of the pulse is only slightly damped (~ 45%) considering the macroscopic distance of 15 cm. From the time delay between the two pressure changes and the known distance, the propagation velocity can be determined. **(b)** Mean values and standard deviations of pulse velocities of five independent measurements as a function of the lateral pressure $\pi$ of the monolayer at 20°C. Around 20 mN/m the velocity possesses a distinct minimum. This corresponds to the phase transition pressure of the membrane, which is indicated by the maximal isothermal compressibility $\kappa_T$ of the interface.

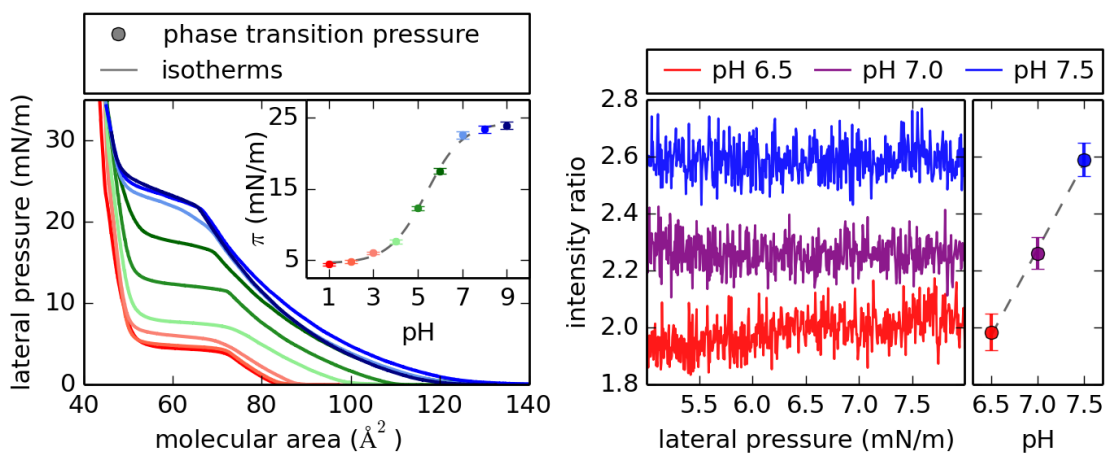

**Fig. 3: (a)** Lateral pressure – area isotherms for varying subphase pH conditions. The plateau regime of the isotherms corresponds to the first order phase transition of the DMPS monolayer. The phase transition pressure $\pi_T$ increases monotonically in a sigmoidal shape for increasing pH bulk values (See inset). This behavior is typical for the $pK_a$-value of the lipid head group. From the first derivate of the sigmoidal fit, we obtain a $pK_a$ of around 5.4 which - in good agreement with literature- corresponds to the $pK_a$ of the carboxyl group of the lipid. **(b)** Intensity ratio $I_R = I_{535nm}/I_{605nm}$ of the fluorescent dye Oregon Green 488 as a function of lateral pressure at different buffer pH-values (6.5, 7 and 7.5). While $I_R$ varies tremendously with bulk pH, its dependence on lateral pressure is negligible (at least for isothermal compression). Plotting mean $I_R$'s for various bulk pH, reveals a linear

relationship in the relevant pressure regime (5 mN/m – 8mN/m). This experimental observation enables us to directly translate $I_R$ measurements into local pH changes under dynamic conditions, i.e. during the propagation of acoustic pulses.

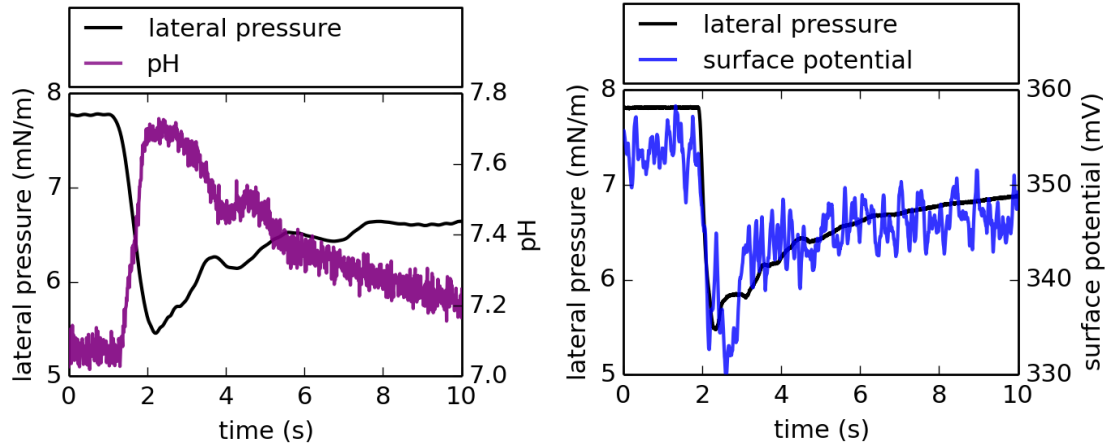

**Fig. 4:** Chemical and electrical properties of the acoustic wave: **(a)** Time course of lateral pressure and interfacial pH for a propagating pulse. Clearly, the two detected signals correlate and the pH rises ~ 0.6 units for the given pressure variation. This is the first observation of pH pulses travelling with the speed of sound along the interface.

**(b)** Readouts of surface potential and lateral pressure during a traveling pulse. The variation of the mechanical signal (~ 2.4 mN/m) coincides with the variation of the electrical response (~ 25 mV), elucidating the coupling of all thermodynamic variables even under dynamic conditions.

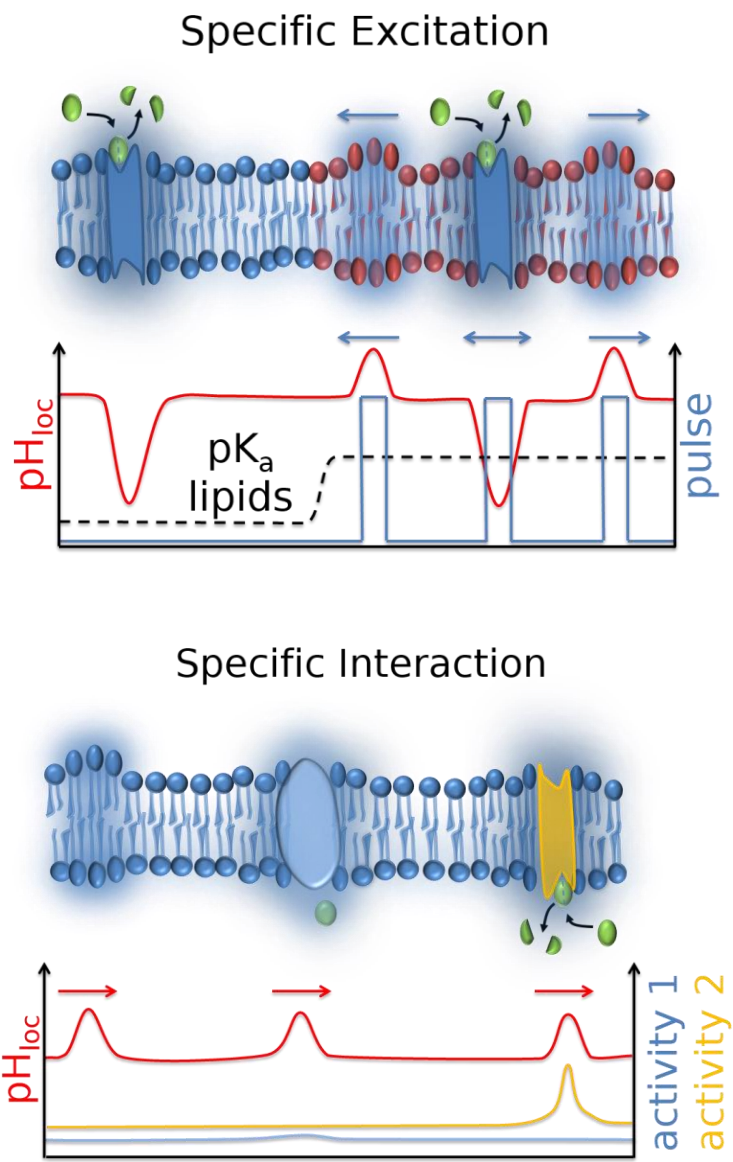

**Fig. 5:** Specific acoustic communication at biological interfaces: **(a)** Specific pulse excitation: An enzyme creates a local change in proton concentration. Only if the resulting $pH_{loc} \sim pK_a$ of the lipids a pulse is excited. **(b)** Specific protein interaction: Enzymes exhibit a maximum activity at a certain pH ($pH_{opt}$). Only if the propagating pH pulse varies the enzyme environment into or out of the $pH_{opt}$ regime, pulse-enzyme interaction by pH is observed and the enzyme can be either "switched-on" or "off".

The interplay between specific excitation depending on the $pK_a$ of the interface and the specific interaction depending on the $pH_{opt}$ of the enzyme results in specific signaling between two enzymes. Of course, coupling of pulses to proteins can also take place electrically via charged groups or mechanically via changes in lipid state and is expected to be particularly increased near the lipid phase transition.

# Supporting Information

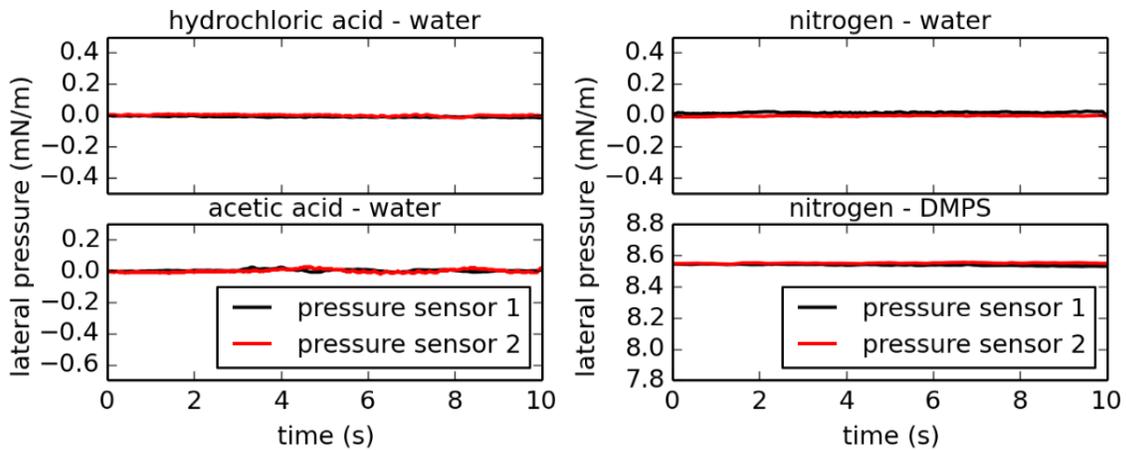

**S1:** Reference measurement on a pure water surface. None of the excitations (hydrochloric acid, acetic acid and nitrogen) are able to evoke any measurable pressure change and hence propagating sound waves on the pure water surface (excitation at ~3 s). Moreover nitrogen does not possess any excitatory effect on DMPS monolayers, too. Consequently all excited pulses in our monolayer experiments have to be due to the interaction between the lipids and the acids (protons).

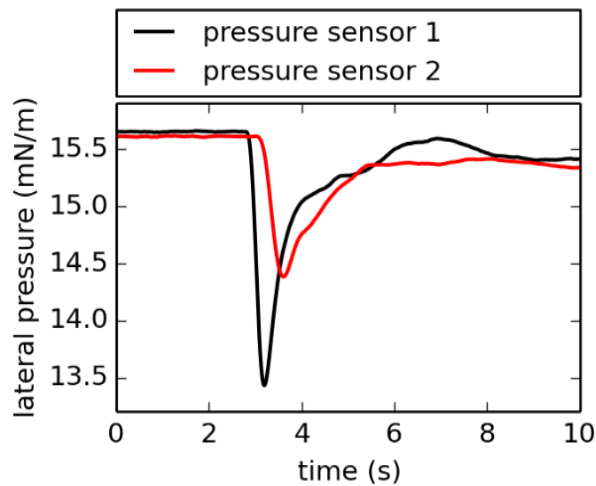

**S2:** Excitation of a propagative pulse across a DMPS monolayer by acetic acid. The time course of the pulse is strikingly similar to the hydrochloric excitation [see Figure 2(a)]. The only common denominator between these two different acids are their protons, elucidating the protonic nature of the excitation.

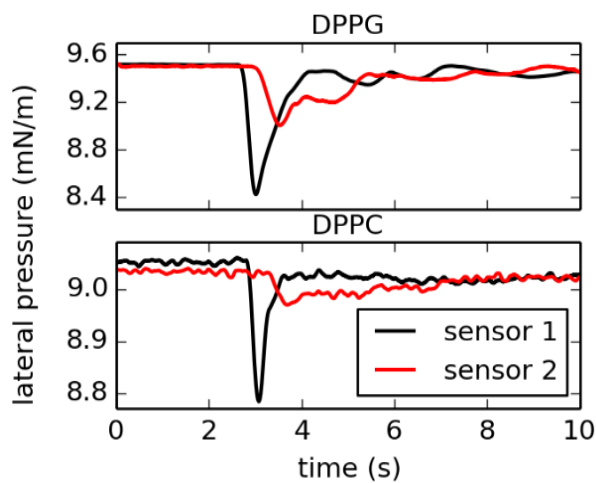

**S3:** Time course of the excitations of DPPC and DPPG monolayers by hydrochloric acid. Surprisingly, pulses can be induced in uncharged (zwitterionic) lipid monolayers like DPPC, too, although with a much smaller amplitude than in charged lipids like DPPG. These experiments illustrate the universality of the pH-driven pulses in lipid systems.

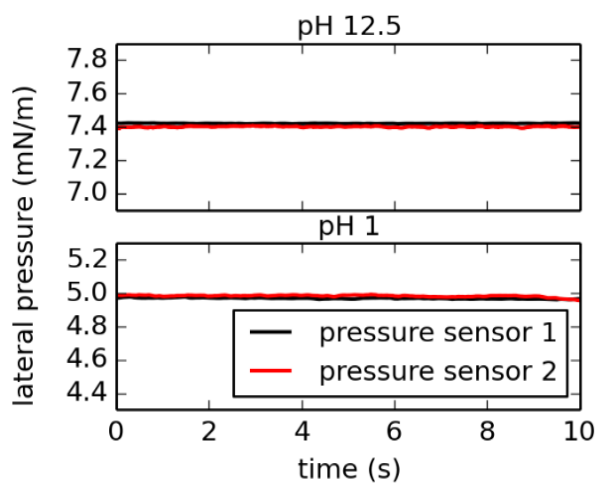

**S4:** At very high, respectively low pH-values (pH 12.5, pH 1), the DMPS monolayer cannot be excited by HCl (excitation at ~3 s). The reason is the $pK_a$-profile of the serine group [see Figure 3(a)]. At low pH-values the lipids are already protonated, while at high pH-values the excitation has to be strong enough in order to facilitate the protonation of the fully deprotonated head groups (25°C, 100 mM NaCl, 10 mM phosphate buffer).

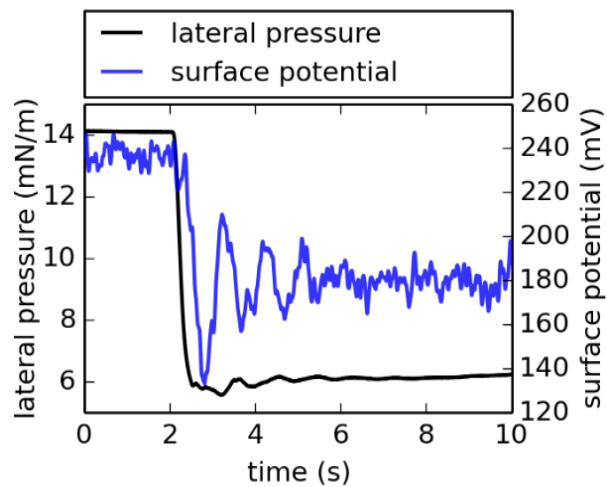

**S5:** For strong excitations and/or less buffer concentrations, pulses with surface potential amplitudes >100 mV were observed (25°C, 10 mM NaCl, 1 mM phosphate buffer). This leads to electric field variation during a pulse of the order of $10^5$ V/m across the surface.